# Coronal temperature profiles obtained from kinetic models and from coronal brightness measurements obtained during solar eclipses

Pierrard V.[1,2], K. Borremans1 and J. F. Lemaire[1,3]
1 Belgian Institute for Space Aeronomy, Space Physics, 3 av. Circulaire, B-1180 Brussels, Belgium.
2 Université Catholique de Louvain, Georges Lemaître Centre for Earth and Climate Research (TECLIM), Place Louis Pasteur 3 bte L4.03.08, 1348 Louvain-La-Neuve, Belgium
3 Université Catholique de Louvain, Center for Space Radiations, 2 chemin du Cyclotron, 1348 Louvain-La-Neuve, Belgium
E-mails: Viviane.pierrard@oma.be, kris.borremans@oma.be; joseph.lemaire@uclouvain.be

**Abstract**
Coronal density, temperature and heat flux distributions for the equatorial and polar corona have been deduced by Lemaire [2012] from Saito's model of averaged coronal white light (WL) brightness and polarization observations. They are compared with those determined from a kinetic collisionless/exospheric model of the solar corona. This comparison indicates rather similar distributions at large radial distances (> 7 Rs) in the collisionless region. However, rather important differences are found close to the Sun in the acceleration region of the solar wind. The exospheric heat flux is directed away from the Sun, while that inferred from all WL coronal observations is in the opposite direction, i.e., conducting heat from the inner corona toward the chromosphere. This could indicate that the source of coronal heating rate extends up into the inner corona where it maximizes at r > 1.5 Rs well above the transition region.

**Introduction**

The base of the corona is a crucial region where the plasma is heated to over 1 MK, where its electron density is known to decrease with large scale height, where the thermal plasma becomes collisionless, where the radial temperature distributions reach a maximum value, and where the solar wind is accelerated to supersonic bulk velocities.

In the present work, the physical properties of the coronal and solar wind expansion have been simulated using steady state hydrodynamical and kinetic models which have been recently reviewed by Echim et al. [2010]. The development of the successive generations of kinetic models (exospheric and Fokker-Planck models) have also been outlined in chronological order by Lemaire [2010]. Pierrard [2012] reviewed the most recent developments with the inclusion of whistler turbulence and kinetic Alfvén waves.

We first calculate the distributions of macroscopic coronal plasma properties (density, expansion bulk velocity, electron and proton temperatures, heat flux, polarization electric field...) by using a Lorentzian exospheric model introduced by Pierrard and Lemaire [1996], and whose exobase boundary conditions (: at 1.07 Rs) are fitted to match typical spectroscopic observations at the base of the corona.

Some of these theoretical predictions are then compared to corresponding plasma properties (density, expansion bulk velocity, electron and proton temperatures, heat flux...) inferred from white light eclipse observations compiled by Saito [1970]. The "dynamical equilibrium method" labeled "dyn" by Lemaire [2012] is used here to determine the experimental radial distributions of these same physical quantities. The latter results from the hybrid model calculation are eventually compared to those of the theoretical/exospheric model.

In regions lower than 0.3 AU, no in-situ observations are available but they can possibly be obtained with future missions going closer to the Sun like Solar Orbiter and Solar Probe. In-situ solar wind observations closest to the Sun were obtained by Helios [Schwenn and Marsch, 1991]. Radioelectric and spectroscopic observations give some information about coronal densities and temperatures of the electrons and ions at very low radial distances (<1.2 Rs) [Esser et al., 1999]. During solar eclipses White Light observations provide also key information on the electron density and temperature profiles in the solar corona up to 10 Rs [Allen, 1947].

Recently, Lemaire [2012] reviewed the capability of standard White Light coronal observations to infer not only radial electron density distributions in the solar corona, but also the radial distribution of electron temperatures. To evaluate this temperature up to 10 solar radii, he introduced a new hybrid method taking into account the continuous expansion of solar corona with supersonic velocities at large radial distances.

By comparing the results obtained by both different approaches (exospheric and hybrid models), we outline and discuss their basic differences in order to take them into account in future modeling efforts and to determine the critical observations that will allow us to identify the physical mechanisms that are relevant in the solar corona and in the solar wind.

**Solar eclipse observations**

It is known since 1941 that electron density profiles can be inferred from White Light (WL) brightness measurements of the corona during solar eclipses [Alfvén, 1941]. Lemaire [2012] presented recently a novel method to determine coronal electron temperatures from these experimental coronal electron density profiles. He labeled it the "/dyn method/" since he considers that the corona is not in hydrostatic equilibrium, but that it is extending beyond eclipse measurements, and expanding up to supersonic velocities as measured at 1 AU. The temperature distributions obtained using this hybrid /dyn method/ are regular solutions of the plasma transport equations, i.e., those used in hydrodynamic solar wind models, but also those of the various generations of kinetic models (exospheric and Fokker-Planck solar wind models) reviewed by Echim et al. [2010]. The hybrid temperature profiles obtained by Lemaire's /dyn method / depend on the radial distribution of the expansion velocity which can be directly determined from the experimental coronal density profile, and the mass flow continuity equation.

We apply the /dyn method/ to Saito's [1970] density model obtained from a compendium of solar eclipse observations at all heliographic latitudes. The latter corresponds to the first empirical two-dimensional (2-D) distribution of average electron densities as a function not only of radial distance, r, but also of heliographic latitude. Saito's empirical 2-D model is based on White Light eclipse observations during minimum solar activity conditions. Equatorial and polar density profiles provided by Saito's model have been used below. They are then compared to density profile corresponding to the exospheric model described in a following section.

The outward bulk velocity of the electrons, $u_e$, used in the /dyn method/ is equal to the bulk velocities of ions species, $u_i$ (since there is no net radial electric current, nor diffusion velocity of particle species with respect to each other). The coronal plasma is treated here as a neutral fluid whose bulk velocities, $u(r) = u_e(r) = u_i(r)$, satisfies the standard mass and momentum flux conservation equations applicable not only to hydrodynamical models, but also to all kinetic solar wind models.

The values of $u_e$ and $n_e$ at 1 AU are free input parameters taken from in-situ measurements at the orbit of Earth. For the equatorial corona and solar wind, we choose: $n_e$ = 5.65 electrons/cm$^3$ and $u_e$ = 329 km/s. For the polar corona and solar wind where fast streams are generally observed, $n_e$ = 2.12 electrons/cm$^3$ and $u_e$ =745 km/s were arbitrarily adopted.

**Exospheric model**s

In low density plasmas like the solar corona and the solar wind, kinetic processes prevail. Exospheric models have been developed for collisionless plasmas, assuming that there are no collisions above a certain radial distance called the exobase. In the equatorial region the exobase for the electrons is typically located at a radial distance smaller than 5 Rs [Lemaire and Scherer, 1970, 1973].

In exospheric models the orbits of particles are affected only by the gravitational force, the electric force and the Lorentz force due to the presence of the interplanetary magnetic field which is assumed here to be radial. Note that spiral magnetic field distributions have also been considered by Pierrard et al. [2001a], but will be ignored here for simplicity, since this complication is not essential below 10 Rs.

The third generation of exospheric models introduced by Pierrard and Lemaire [1996] and extended by Maksimovic et al. [1997] as well as by Lamy et al. [2003] will be used in this study. Indeed, a Lorentzian or "Kappa" velocity distribution (VDF) has been used at the exobase for the electrons and a Maxwellian VDF for the protons. The Kappa VDFs decrease as a power law of the energy so that the value of the kappa index determines the slope of the energy spectrum of the suprathermal electrons forming the tail of the VDF. In the limit $\kappa \to \infty$, the Kappa function degenerates into a Maxwellian. The Liouville equation is solved analytically and numerically to determine the velocity distribution function of the electrons and protons at larger radial distances. Once the VDFs are obtained, the densities of electrons and protons have been calculated, and the electrostatic potential distribution has been determined to satisfy the quasi-neutrality condition of the plasma. Finally, all other higher order moments of the VDFs have been calculated. They were then compared to those derived for the hybrid models which were obtained by Lemaire's /dyn method/ from Saito's equatorial and polar electron density profiles.

A Kappa distribution is chosen for the electrons since in-situ VDFs measured at 1 AU in the solar wind are characterized by a thermal core population and a halo of suprathermal electrons [Pierrard et al., 2001b]. Such distributions with suprathermal tails are conveniently fitted by Kappa or Lorentzian VDFs. The evidence of such power law velocity distributions in the case of many observed space plasmas suggests the existence of a universal mechanism generating VDFs with such suprathermal tails [Pierrard and Lazar, 2010].

The calculated radial distributions of the lower order moments of the electron VDFs are illustrated by the solid lines in Fig. 1. These radial profiles were obtained by assuming that the exobase level is at an altitude of 0.07 Rs, (r = 1.07 Rs), that the temperatures of the electrons and protons at the exobase are $T_e = T_p = 8 \ 10^5$ K, and their densities $n_e = n_p = 10^8$ m$^{-3}$.

**Comparison of the theoretical and experimental results**

Figure 1 illustrates the profiles obtained between altitudes from 0.07 Rs to 10 Rs with the two different approaches (exospheric and hybrid models). The exospheric profiles correspond to the solid line. Those corresponding to Saito's equatorial empirical density profile are shown by the dashed-dotted line. Those corresponding to Saito's polar region are illustrated by dashed lines.

The kappa parameter for the exospheric model has been chosen to be 3.5 to obtain the same bulk velocities at 1 AU as the Saito polar model.

Panel 1 illustrates the number density profiles of the electrons which is necessarily equal to that of the protons since no other ion species in considered in our model calculations. It can be seen that the exospheric density decreases more slowly as a function of the altitude than Saito's equatorial and polar density distributions. Note that this conclusion remained even when other values are adopted for the index kappa of the electron VDF. The empirical number density observed in the coronal polar region (dashed) is lower and decreases faster than in the equatorial region (dashed-dotted), as indeed observed in coronal holes in comparison with the equatorial streamers.

Panel 2 illustrates the electric potential obtained in the exospheric model. This is a crucial physical quantity in solar wind models since it determines the electrostatic force that accelerates the protons (and other ions) to supersonic bulk speeds at large radial distances. This was first emphasized by Lemaire and Scherer [1970, 1973]. See also Lemaire and Pierrard [2001] for a modern description of this acceleration mechanism.

Note that the electrostatic potential difference between the exobase and the interplanetary medium is enhanced when the population of suprathermal electrons is arbitrarily enhanced by decreasing the value of the kappa index [Pierrard and Lemaire, 1996; Maksimovic et al., 2001]. In the model illustrated in Fig. 1, kappa=3.5 and the exobase electrostatic potential is equal to +2600 Volts.

Panel 3 represents the escape flux of the coronal electrons which have a sufficient large outward velocity to overcome the electrostatic potential energy. This flux is of course equal to the outward flux of the protons accelerated out of the corona by this polarization electrostatic field.

Panel 4 shows the bulk velocity of the electrons as a function of altitude. It is of course equal to the bulk velocity of the protons since the net electric current is postulated to be equal to zero and because no other ion species are assumed to be present in our simple exospheric model. Note that both bulk velocities are directed away from the Sun and increase as a function of the altitude just as in Parker's well-known hydrodynamic model. The region of steepest acceleration occurs to be at a radial distance of about 2 Rs in all three models. Note that the bulk velocity at 1 AU (not shown because out of scale) was chosen to be the same for the exospheric model and Saito's polar one. It is interesting to point out that in the hybrid model based on Saito's polar density distribution the coronal plasma gets accelerated more strongly and at altitudes closer to Sun than in our theoretical model.
With a higher value of kappa (less suprathermal electrons), the calculated exospheric bulk velocity is reduced and the bulk velocity profile becomes closer to that of Saito's equatorial model (not shown for clarity of the plots).
It can also be seen that over the equatorial region (dashed-dotted curve) the acceleration of the coronal plasma is smoother (not so steep) than over the pole and presumably in coronal holes (dashed curve).

Panel 5 illustrates the theoretical proton temperatures: the total or pitch angle averaged temperature (dotted line); the parallel temperature (upper line at 10 Rs); the perpendicular temperature (bottom line at 10 Rs). Note that the theoretical perpendicular proton temperature has a maximum at a radial distance of 2 Rs corresponding to the acceleration region; beyond this altitude it decreases rather steeply as $1/r^2$, i.e., proportionally to B(r) due to the adiabatic invariance of the magnetic moment of the ions. It can be seen that $T_{perp} > T_{//}$ at r < 2 Rs in the exospheric model, while $T_{perp} < T_{//}$ at larger radial distances [Lemaire and Pierrard, 2001]. Such different temperature anisotropies are indeed observed in the inner solar corona by EUV spectroscopic observations [Cranmer, 2002 for a review of the coronal holes characteristics], as well as at large radial distances by in situ solar wind measurements [Schwenn and Marsch, 1991].

The proton temperature deduced from Saito empirical density distributions (not illustrated) is arbitrarily assumed to be isotropic, furthermore it is assumed to be equal to the electron temperature at all radial distances. In future more sophisticated version of this hybrid (/dyn/) model, the effect of temperature anisotropies can in principle be taken into account, of course. In the current version of Lemaire's hybrid model the proton temperature distribution is assumed to have the same distribution as the electron temperature. Note that the latter is illustrated in panel 6 and has a maximum value at the same altitude as the /dyn/ electron temperature distribution.

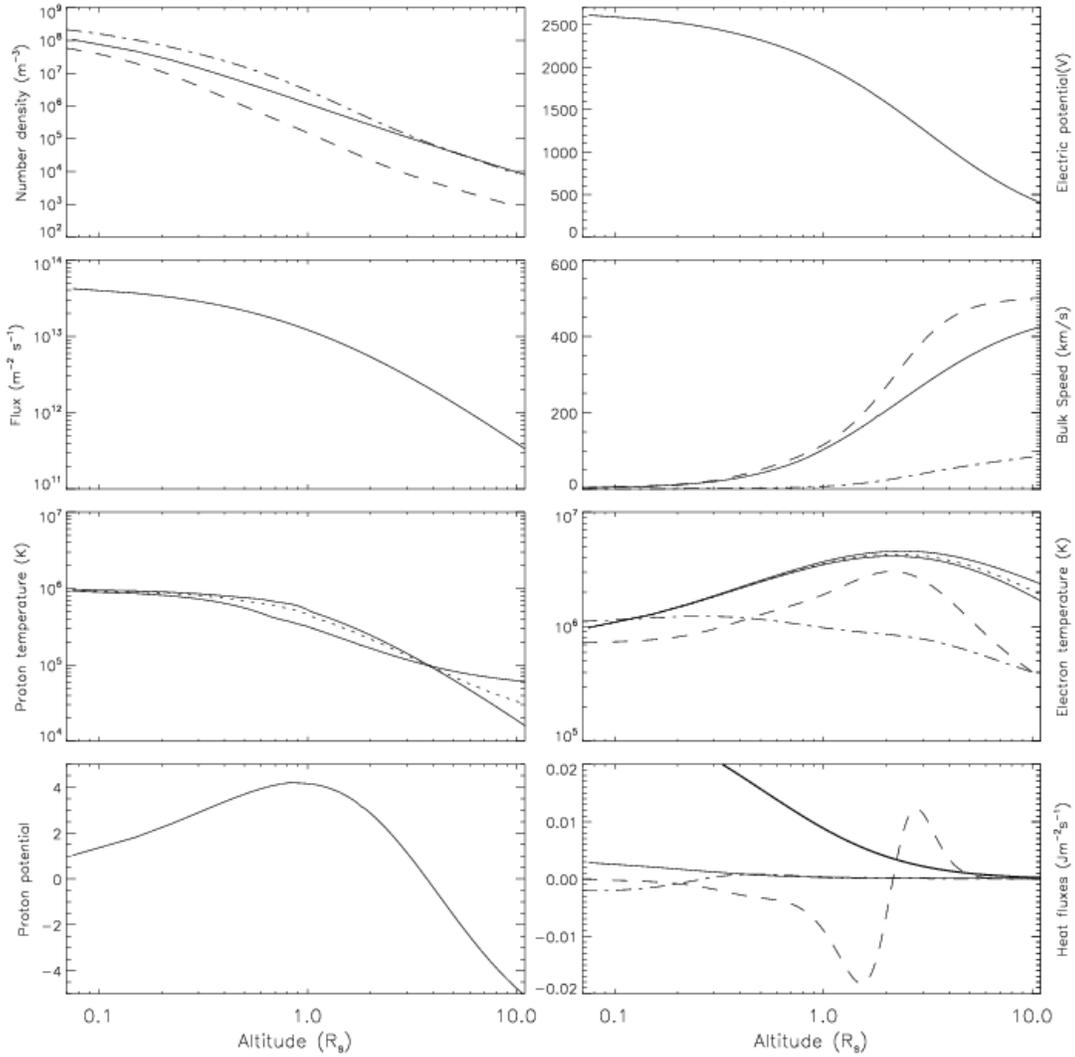

Fig. 1: Profiles obtained with the exospheric model (solid line), from Saito's equatorial density profile (dashed-dotted line), and from Saito's empirical density profile over the polar regions (dashed line). Panel 1: number density of the electrons; Panel 2: polarization electric potential satisfying the quasi-neutrality of the exospheric coronal plasma; Panel 3: distribution of the theoretical electron flux; Panel 4: plasma expansion velocities; Panel 5: exospheric proton temperatures (total or pitch angle averaged: dotted line; parallel temperature: upper curve at 10 Rs; perpendicular temperature: bottom curve at 10 Rs); Panel 6: electron temperatures (total exospheric temperature: dotted line; parallel exospheric temperature: upper solid line at 10 Rs; perpendicular exospheric temperature: bottom solid line at 10 Rs; Saito equatorial: dashed-dotted line; Saito polar: dashed line); Panel 7: normalized total proton potential energy distribution in the exospheric model; Panel 8: radial heat fluxes carried by the electrons in the exospheric model:

bold upper solid line; carried by the protons in the exospheric model: thin solid line; in the hybrid model based on Saito's empirical polar density: dashed line; in hybrid model based on Saito equatorial: dashed-dotted line).

Recent SOHO observations show that the temperatures are different for the different ion species and for the electrons at radial distance as low as 2 Rs [Esser and Edgar, 2000]. This indicates that further improvement of the hybrid model would have to be developed, but this will require additional observational inputs: e.g. empirical radial distributions of the coronal ion temperatures and of their anisotropies.

Panel 6 illustrates the electron temperature profiles of the exospheric and hybrid models. The exospheric model gives the highest temperatures (parallel temperature: upper solid line; perpendicular temperature: bottom solid line; pitch-angle averaged temperature: dotted line). The theoretical electron temperatures have a maximum at an altitude of 2 Rs. By chance this corresponds almost where Saito's polar temperature profile (dashed line) has also its maximum value. The peak of temperature is however spread over a wider range of altitudes for the exospheric model than for the empirical temperature profile over the polar region. Note that Saito's hybrid equatorial temperature (dashed-dotted line) has a smaller maximum value, and that the latter is located closer to the base of the solar corona, at an altitude of $h_{max}$ = 0.2 Rs ($r_{max}$ = 1.2 Rs).

It is noteworthy that at the base of the corona the electron temperatures and densities of the hybrid models are smaller over the polar region (coronal holes) than over the equatorial region. These results are consistent with all those obtained by Lemaire [2012] from other eclipse WL observations with his /dyn method/. These results are also in qualitative agreement with the densities and temperatures deduced otherwise from coronal holes UV and X-ray observations made on board of Skylab, and other modern space missions (see Cramner [2009] for a review).

The larger the value of the kappa-index, the lower is the radial distance where the theoretical electron temperature reaches its maximum value. At large radial distances, the temperature decreases as predicted in all earlier kinetic models. What was not apprehended before the development of the third generation of exospheric models by Lamy et al. [2003] was that by lowering the exobase altitude the theoretical temperatures have a peak in the middle corona and that below this maximum it decreases with altitude to possibly match with the chromospheric temperatures. Therefore, a theoretical maximum temperature can be produced in exospheric models when their exosphere is arbitrarily assumed to start at smaller radial distances, i.e., by lowering their exobase altitude.

Here, we have assumed the exobase to be at r = 1.07 Rs. Of course at such low altitudes the plasma is generally collision-dominated, and, in principle, exospheric models fail to be appropriate due to the large Knudsen number of the coronal plasma at these low altitudes! Nevertheless, Meyer-Vernet [2006] showed that even at low altitudes, the plasma cannot be considered as dominated by collisions, due to the long range properties of the Coulomb interaction. Indeed, since the particle free path increases as $v^4$ in a plasma, the suprathermal particles are non-collisional while thermal and sub-thermal particles experience many more collisions per unit time and thus have much smaller free pathes. This does justify that in the solar transition region, the corona and the wind, the heat flux is not given by classical Spitzer and Harm expression [Shoub, 1983].

Panel 7 illustrates the normalized proton total potential (gravitational + electrostatic potentials) in the case of the exospheric model. It has been verified at $h_{max}$ < 1 Rs (i.e. below the maximum of the proton potential) the electrostatic force acting on a proton is smaller than the gravitational force. As a matter of consequence, at these lower coronal altitudes, protons with kinetic energies smaller than the maximum total potential are unable to evaporate and escape to infinity into interplanetary space. Only protons with large enough kinetic energies contribute to the solar wind particle flow and to their outward energy flux.

At $h_{max}$ ~ 1 Rs ($r_{max}$ ~ 2 Rs) the outward electric force (eE) acting on protons balances the inward gravitational force (mg). It is only beyond this distance that protons are effectively accelerated out of the corona to supersonic velocities. Below this altitude, their kinetic energy and velocity is decreasing with r. This is a key result that was first pointed out by Lemaire and Scherer [1970, 1973] and often overlooked within the hydrodynamic modeling community.

The solid lines in Panel 8 illustrate the heat fluxes defined (as usually) in the frame of reference co-moving with the bulk velocity of the expanding plasma. The exospheric heat flux carried by the electrons is shown by the bold upper solid line; the theoretical heat flux transported away from the Sun by the exospheric protons is given by the thin solid line; the heat flux transported by the electrons in Saito's empirical polar density model is given by the dashed line; the same heat flux distribution in Saito's empirical equatorial density model is given by the dashed-dotted line.

Note that in the exospheric model, the heat fluxes carried by the electrons (upper solid curve) and by the protons (lower solid curve) are everywhere positive. They are everywhere decreasing functions of h and r. At large distances, the theoretical values tend asymptotically to zero as in the hybrid models associated with Saito's density profiles.

From the Panel 8 it can be inferred that close to the base of the corona the outward directed exospheric energy fluxes increase dramatically with decreasing altitudes. However, the heat conduction fluxes obtained in the hybrid models consistently change sign where the corresponding electron temperature gradients reverse from negative values (beyond $h_{max}$) to positive values (below $h_{max}$). The challenge is then to identify what could be the most relevant sources of energy that can account for such distributions of the coronal heat fluxes and of the required coronal heating rates which result in electron temperature profiles like those predicted by Lemaire's [2012] hybrid models in the equatorial and polar regions of the solar corona.

Here appears a major difference between exospheric and hydrodynamic models of the solar corona. In exospheric models, collisions are ignored above an arbitrarily fixed exobase altitude. No extra coronal heating and energy deposition is then considered. The temperature increases naturally at low radial distances in the solar corona due to the so-called velocity filtration effect [Scudder, 1992]. In this case, the heating of the corona is suprathermally driven, i.e., it may be explained simply by the presence deep in the corona of a population of suprathermal particles, without the need of an additional or ad-hoc energy source [Pierrard and Lamy, 2003]. This tentative speculation and its observational investigations are discussed in Pierrard and Lazar [2010].

**Conclusions**

The */dyn method/* developed by Lemaire [2012] has been used with Saito's average electron density models which were derived from coronal WL eclipse observations. This method is a new diagnostic tool for the determination of the electron temperatures distributions in the solar corona, where their maximum in the equatorial and polar corona. It is therefore a diagnostic tool to determine where the coronal heating rate maximizes in these coronal regions.

On the other hand, an exospheric solar wind model of third generation has been used to determine the distributions of the coronal electron density, bulk velocity, of the electron and proton temperatures and of the heat flux. In this Lorentzian exospheric model there are no collisions above an exobase radial distance arbitrarily chosen to be at $r = 1.07$ Rs. No extra heating rate above this altitude is of course assumed as in exospheric SW models.

The theoretical electron densities, temperatures and heat flux distributions corresponding to this exospheric model have then be compared with those of the Lemaire's hybrid model based on Saito's empirical density distributions for the equatorial and polar regions of the solar corona.

Exospheric models with arbitrary low exobase altitudes exhibit a similar temperature maximum above the transition region as the hybrid models, although the physical mechanisms accounting for this temperature maximum are quite different. This leads to very different heat flux profiles obtained in the exospheric model and in the hybrid models. We conclude thus that the distribution of heat flux should be viewed as a key proxy to identify what physical mechanisms are most relevant in determining the thermal structure of the solar corona.

**Acknowledgments**
The research leading to these results has received funding from the European Commission's Seventh Framework Program (FP7/2007-2013) inside the grant agreement SWIFF (project n°2633430, www.swiff.eu). This project was also supported by IUAP (Interuniversitary project) CHARM. V. Pierrard thanks also the STCE (Solar-Terrestrial Center of Excellence) and BISA for their support.